\documentstyle[12pt, epsf]{article}

\advance\hoffset -0.5cm
\advance\voffset -1.0cm
\newlength{\height}
\divide \height by 3

\begin{document}


\baselineskip=\height


\begin{titlepage}
\begin{center}

\makebox[\textwidth][r]{SNUTP 95-114}

\vskip 0.35in
{{\Large \bf Anomalous Magnetic Moment of Anyons \\
in three dimensional $CP^{N-1}$ Model}}
\end{center}
\begin{center}
\par \vskip .1in \noindent Deog Ki Hong\footnote[1]
{E-mail address: dkhong@hyowon.cc.pusan.ac.kr}
\end{center}
\begin{center}
Department of Physics, Pusan National University\\
Pusan 609-735, Korea\\

\begin{center}
\par \vskip .1in \noindent Jin Young Kim\footnote[2]
{E-mail address: jkim@kowon.dongseo.ac.kr}
\end{center}
\begin{center}
Division of Basic Sciences, Dongseo University\\
Pusan 616-010, Korea\\
\end{center}

\par \vskip .1in
\noindent
\end{center}

\begin{abstract}
We calculate the anomalous magnetic moment of anyons
in three dimensional $CP^{N-1}$ model with a
Chern-Simons term in various limits in $1/N$ expansion.
We have found that for anyons of infinite mass
the gyromagnetic ratio ($g$-factor) is 2 up to the next-to-leading
order in $1/N$.
Our result supports a recent claim that
the $g$-factor of nonrelativistic anyons is
exactly two. We also found that
for $-{8\over\pi }<\theta<0$, the electromagnetic
interation between two identical aynons
of large mass are attractive.

\vskip 0.2in
\noindent


\end{abstract}

\end{titlepage}

One of the peculiarities of planar physics is that
there exists a particle of arbitrary spin, called anyon, since
the rotation group in the two-dimensional space is
$U(1)$.
In field theories, anyon can be realized by minimally coupling
a boson or a fermion to Chern-Simons gauge fields\cite{zee}.
Due to the coupling to the Chern-Simons gauge fields,
the wave function of two identical particles
gets an Aharonov-Bohm phase, $4\pi\over\kappa$,
where $\kappa$ is the coefficient of the Chern-Simons term in the
Lagrangian.
Since the Aharonov-Bohm phase is interpreted as a
statistical phase,  the particle gets an induced spin,
$S=1/\kappa$.
Anyon has been studied extensively because of its possible
application to some of the planar
condensed matter systems. It is found that
anyon matter exhibits superconductivity\cite{chen}
and describes quantum Hall system\cite{arovas}.

To study the many-body effects of anyons,
one needs to find out how they interact with each other.
Besides the statistical interaction due to
the Chern-Simons fields, anyons may interact
electro-magnetically.
Especially, the interaction between its magnetic moment and
electric charge can be very important since it may lead to
a bound state of anyon and a magnetic flux tube\cite{kogan},
which is introduced by Jain\cite{jain} to describe fractional
quantum Hall effect. Depending on the sign of the
magnetic moment of anyon, the interaction of identical anyons
could be attractive
while the electric charge-charge interaction is always repulsive.
Therefore, for a certain value of the coefficient of
the Chern-Simons term, two identical anyons can form a bound state.
Finding out the exact value of the magnetic moment of anyon is
therefore of some importance.

There have been several calculations on the magnetic moment of anyon.
Chou {\it et al.} showed that the $g$-factor of
nonrelativistic anyonic particles is exactly two \cite{nair}.
Also, in field theoretical models, it is shown  that
in the pure anyonic limit
(namely for large value of the coefficient of the Chern-Simons term)
the gyromagnetic ratio of anyon at one-loop
is $g=2+O\left({1\over\kappa}\right)$
where $\kappa$ is the coefficient of the Chern-Simons
term\cite{semenoff, gat}.

In this letter, we calculate
the magnetic moment of anyons in
the (2+1)-dimensional $CP^{N-1}$ model
with a Chern-Simons term\cite{park, rajeev} and we try to
see if we reproduce the Chou {\it et al.}'s result for anyons
of large mass.
For $N=2$, the model was originally
studied to describe the high $T_c$ superconductivity
\cite{polyakov, wiegman}, since the model appears as a
long-wavelength limit of the Hubbard model
near half-filling\cite{wiegman}.

The $CP^{N-1}$ model with a Chern-Simons term is described by a
Lagrangian density (in the Euclidean notation)
\begin{equation}
{\cal L}_{CP^{N-1}}=|D_{\mu}n_i|^2+
      \alpha\left(|n_i|^2-{N\over g^2}\right)
      +iN{\theta\over 16}\epsilon_{\mu\nu\lambda}
              A_{\mu}\partial_{\nu}A_{\lambda},
\label{lagrangian}
\end{equation}
where $i=1,\cdots,N$ and
the covariant derivative $D_{\mu}=\partial_{\mu}+iA_{\mu}$.
In the original $CP^{N-1}$ model without the Chern-Simons term,
the vector field $A_{\mu}$ is an auxiliarly field satisfying
the constraint
$A_{\mu}=(i/2){\bar n_i}{\stackrel{\leftrightarrow}
\partial_{\mu}}n_i$
which reduces the independent components of the complex fields
$n_i$ by 2 together with the constraint on the ``length square"
of $n_i$, $|n_i|^2={N\over g^2}$. But, as is well known,
the quantum corrections generate a kinetic term for $A_{\mu}$.
With the Chern-Simons term
the $CP^{N-1}$ model behaves quite differently from the
original one. The added Chern-Simons term in the lagrangian
(\ref{lagrangian}) not only makes the vector
field $A_{\mu}$ propagte even at the classical level but also
provides an induced fractional spin to $n_i$
by the Aharonov-Bohm effect \cite{aharonov}.

In the $1/N$ perturbation, the model exhibits a nontrivial phase
structure\cite{arefyeva, park, rajeev}.
For ${1\over g^2}<{1\over g_c^2}\equiv\int [d^3p/(2\pi)^3](1/p^2)$,
$\left<\alpha \right>\ne0$ and $\left< n_i\right>=0$.
Therefore, the $n$ fields get mass,
$M=\sqrt{\left<\alpha\right>}$, and the flavor
symmetry $SU(N)$ is unbroken. For ${1\over g^2}>{1\over g_c^2}$, the
flavor symmetry breaks down to $U(N-1)$, leaving $2N-2$ massless
Nambu-Goldstone  bosons.

We calculate the anomalous magnetic moment of anyons
described by (\ref{lagrangian}) in
the symmetric phase.
It is well known that the magnetic moment of a particle can be
infered from the on-shell
matrix element of the effective interaction Hamiltonian under
a (spatially) uniform external magnetic field;
\begin{equation}
\left<p^{\prime}|{\cal H}_{\rm int}|p\right>=
-\int d^2xA_{\mu}(x) \left<p^{\prime}|j^{\mu}(x)|p \right>,
\label{hamiltonian}
\end{equation}
where $j_{\mu}=i{\bar n_i}{\stackrel{\leftrightarrow}
\partial_{\mu}}n_i$.
The matrix element of the current of a scalar field is given by
\begin{equation}
\left<p^{\prime}|j^{\mu}(x)|p \right>=P^{\mu}F_1(q^2)-{i\over 2M}
\epsilon^{\mu\nu\lambda}q_{\nu}P_{\lambda}F_2(q^2),
\end{equation}
where $P^{\mu}=p^{\prime\mu}+p^{\mu}$ and
      $q^{\mu}=p^{\prime\mu}-p^{\mu}$. $F_1(q^2)$ corresponds
to the electric charge form factor.
For a constant magnetic field, $B$, one can easily see that,
as $p^{\prime}\to p$,
\begin{equation}
\left<p^{\prime}|{\cal H}_{\rm int}|p\right>=- {B\over 2M}F_2(0).
\end{equation}
Therefore $\mu=F_2(0)/2M$ is
the magnetic moment of a scalar field which couples minimally to
photon.
There are seven diagrams, shown in fig.1, which are leading order
corrections in $1/N$ to the vertex function.
By Lorentz invariance, the diagram (b) must be proportional to
$q_{\mu}$, which is then zero due to the current conservation.
The diagrams (c) and (d) are zero by explicit calculations.
The diagram (e) in fig.1 vanishes identically,
because of the symmetry of the Lagrangian under $n^i\to \bar n^i$,
$A_{\mu}\to -A_{\mu}$, which is similar to Furry's theorem.
Finally, one can easily see that the parity-odd parts of
the diagrams (f) and (g) are zero by the similar argument used to
prove the Coleman-Hill theorem on non-renormalization of the
Chern-Simons term \cite{coleman}.
The diagram (f) gives
\begin{equation}
\Gamma_{\mu}^{(\rm f)}=\int {d^3k\over (2\pi)^3}
G_{\alpha\beta}(k){(2p-k)_{\alpha}\over (p-k)^2+M^2}
\Sigma(q-k)\Gamma_{\beta\mu}(k,q-k,-q),
\label{diaf}
\end{equation}
where $G_{\alpha\beta}(k)$ is the propagator for the gauge fields
and $\Sigma(q-k)$ is the propagator for $\alpha$ field.
$\Gamma_{\beta\mu}(k,q-k,-q)$ is the one-loop effective
$AA\alpha$ vertex. Because of the current conservation and the
analyticity of the vertex, $\Gamma_{\beta\mu}(k,q-k,-q)=k_{\beta}
G_{\mu}(k,q)$, where $G_{\mu}(k,q)$ is analytic at the zero momenta.
(Note that the effecive vertex is analytic at the vanishing
external momenta, since the $n$ fields are massive.)
Since in the eq. (\ref{diaf}) $\Gamma_{\beta\mu}(k,q-k,-q)$ is
contracted with the gauge field propagator $G_{\alpha\beta}(k)$,
only the parity-even part of the propagator survives in
the eq. (\ref{diaf}). Therefore the only diagram which has a
nonvanishing parity-odd part is diagram (a). Since the magnetic
moment is parity-odd, it is the only diagram that contributes to the
magnetic moment of anyon in the leading order
in $1/N$.
We calculate the diagram (a) under the condition that both initial
and final $n$ fields are on the mass-shell;
\begin{equation}
\Gamma_{\mu}=\int {d^3k\over (2\pi)^3}
     {(2p-k)_{\lambda}(2p-2q-k)_{\nu}(2p-2k-q)_{\mu} \over
     \left[(p-k)^2-M^2 \right]\left[(p-k-q)^2-M^2\right]}
    G_{\lambda\nu}.
\label{vertex}
\end{equation}
We follow the Feynman rules
for the $CP^{N-1}$ model derived in the reference \cite{park}.
To utilize the Feynman parametrization in evaluating the vertex
function (\ref{vertex}) we rewrite the gauge field propagator
in the
K\"allen-Lehman representation, in which the propagator
is given by
\begin{equation}
  G_{\mu\nu}(p)=\int_0^{\infty}ds{\rho_1(s)\over p^2-s+i\epsilon}
  \left(\delta_{\mu\nu}-{p_{\mu}p_{\nu}\over p^2}\right)+
  {\theta\over16}
  \int_0^{\infty}ds{\rho_2(s)\over p^2-s+i\epsilon}
  \epsilon_{\mu\rho\nu}p^{\rho},
\label{propagator}
\end{equation}
where
\begin{equation}
   \rho_1(s)={i\over\pi}{\Gamma(s)\over \Gamma^2(s)+
  \left({\theta\over16}\right)^2s}, \quad
  \rho_2(s)={i\over\pi}{1\over \Gamma^2(s)+
  \left({\theta\over16}\right)^2s}
\end{equation}
with
\begin{equation}
\Gamma(s)={1\over2}(s+4M^2){\tan^{-1}{\sqrt{s}\over 2M}\over
4\pi\sqrt{s}}-{M\over4\pi}.
\end{equation}

The parity-odd part of the vertex function which contributes to
the magnetic moment is
\begin{equation}
 \Gamma_{\mu}^{\rm odd}={\theta\over16}\int_0^{\infty}ds\rho_2(s)
  \int {d^3k\over (2\pi)^3}
  {(2p-k)_{\lambda}(2p-2q-k)_{\nu}(2p-2k-q)_{\mu}
  \epsilon_{\lambda\rho\nu}k^{\rho}\over
  \left[(p-k)^2-M^2\right]\left[(p-k-q)^2-M^2\right]
  \left[ k^2-s+i\epsilon\right]}.
\label{vertexodd}
\end{equation}
After some calculations, we get
\begin{equation}
 \Gamma_{\mu}^{\rm odd}=-{i\theta\over32\pi}\epsilon_{\lambda\nu\mu}
  p_{\lambda}q_{\nu}\int_0^{\infty}ds\rho_2(s)F(M^2,s,q^2),
\end{equation}
where
\begin{equation}
 F(M^2,s,q^2)=\int_0^1dx\int_0^x dy
 \left[ M^2x^2+(s-i\epsilon)(1-x)-q^2y(x-y)\right]^{-1/2}.
\end{equation}
The magnetic form factor is then
\begin{equation}
F_2(q^2)={-M\theta\over32\pi}\int_0^{\infty}ds\rho_2(s)F(M^2,s,q^2).
\end{equation}
For small momentum transfer ($q^{\mu}\to0$)
\begin{equation}
  F(M^2,s,q^2)={1\over M}-{\sqrt{s}\over M^2}+
  {s\over 2M^3}\ln \left( 1+2\sqrt{M^2/s}\right)+
  O\left( q^2\right),
\end{equation}
we get
\begin{equation}
 \mu=-{i \theta\over64\pi^2} \int_0^{\infty}
   {ds\over \Gamma^2(s)+\left({\theta\over16}\right)^2s}
   \left[ {1\over M}-{\sqrt{s-i\epsilon}\over M^2}+
  {s-i\epsilon\over 2M^3}\ln
\left( 1+2\sqrt{{M^2\over s-i\epsilon}}\right)\right].
\label{moment}
\end{equation}
In the leading order in $1/N$, the eq. (\ref{moment}) is
the exact magnetic moment of $n$ fields
in the $CP^{N-1}$ model with a Chern-Simons term. Since
it is not easy to perform the integration, we consider two
extreme cases where $(\romannumeral 1)$ the mass of
$n$ fields is very large ($M\to\infty$) and $(\romannumeral 2)$
very small ($M\to0$). In these limits the parity-odd part of
Chern-Simons propagator becomes for each cases
\begin{eqnarray}
(\romannumeral 1)\quad  & G^{a(\rm odd)}_{\lambda\nu}(k) \simeq
     {16\over\theta}
  {\epsilon_{\lambda\rho\nu}k^{\rho}\over k^2},\quad (M^2\gg k^2)\\
(\romannumeral 2)\quad & G^{a(\rm odd)}_{\lambda\nu}(k) \simeq
{16\theta\over \theta^2+1}
 {\epsilon_{\lambda\rho\nu}k^{\rho}\over k^2},\quad (M^2\ll k^2).
\label{limits}
\end{eqnarray}
Repeating calculations with these propagators, we get
\begin{eqnarray}
    (\romannumeral 1)\quad & \mu = {4\over \pi M\theta}
             \quad (M\to \infty),\\
     (\romannumeral 2)\quad & \mu = {4\theta\over \pi M(\theta^2 +1)}
              \quad(M\to0).
\end{eqnarray}
Since the induced spin of $n$ fields is $S={4\over\pi\theta}$, the
gyromagnetic ratio is for each limiting cases
\begin{eqnarray}
(\romannumeral 1)\quad & g=2 \quad (M\to \infty), \\
(\romannumeral 2)\quad & g={2\theta^2\over \theta^2+1}\quad(M\to0).
\end{eqnarray}
We see here that in the non-relativistic limit (namely in the heavy
mass limit, $M\to\infty$) the $g$-factor of  anyons
is two at the leading order in the $1/N$ expansion.

We now consider the next-to-leading corrections to the magnetic
moment when the mass of anyon goes to infinity to see if the
$g$-factor of heavy anyon does not get corrections in higher orders in
$1/N$. The next-leading order diagrams in $1/N$ for the vertex
function are shown in fig.2. Since in the  large $M$ limit, the gauge
field propagator becomes
\begin{equation}
G_{\mu\nu}(p)=
{32\over 3\pi\theta^2} {1\over M} \left( \delta_{\mu\nu}-
{p_{\mu}p_{\nu}\over p^2}\right)+
{16\over \theta}{\epsilon_{\mu\rho\nu}p^{\rho}\over p^2},
\end{equation}
the contribution of
the parity-even part of the gauge field propagator is suppressed
by $1/M$. Therefore, in the large $M$ limit, one may neglect
the parity-even part of the gauge field propagator in calculating
the vertex function.
To get the parity-odd vertex correction,
we need to consider the diagrams which has odd number of the gauge
propagator.
Then, the diagram (a) in fig. 2 is
the only diagram that may contribute to
the parity-odd part of the vertex function in the next-to-leading
order in $1/N$. But, one can see that the diagram does not
contribute to the magnetic moment of heavy anyon by the Coleman-Hill
argument we used previously. The diagram (a) gives in the large $M$
limit

\begin{eqnarray}
 \Gamma_{\mu}^{(2)}\!\!\!&=&\!\!\!{1\over N}
\left( {16\over\theta}\right)^3
\int_{k_1,k_2}
{(2p-k_1)_{\delta}(2p-2k_1-k_2)_{\rho}
(2p-q-k_1-k_2)_{\sigma}\over \left[ (p-k_1)^2-M^2\right]
\left[ (p-k_1-k_2)^2-M^2\right]} \nonumber \\
\!\!\!&\cdot&\!\!\!\!\!
{\epsilon_{\delta\nu\alpha}k_1^{\nu}\over k_1^2}
{\epsilon_{\rho\tau\beta}k_2^{\tau}\over k_2^2}
{\epsilon_{\sigma\kappa\gamma}(q-k_1-k_2)^{\kappa}\over
(q-k_1-k_2)^2} \Gamma_{\mu\alpha\beta\gamma}(q,k_1,k_2;q-k_1-k_2),
\label{next}
\end{eqnarray}
where $\Gamma_{\mu\alpha\beta\gamma}(q,k_1,k_2;q-k_1-k_2)$ is the
1-loop 4-photon effective vertex. Because of the gauge
invariance and the analyticity of the effective vertex\cite{coleman},
\begin{equation}
 \Gamma_{\mu\alpha\beta\gamma}(q,k_1,k_2;q-k_1-k_2)=q_{\mu}
{k_1}_{\alpha} {k_2}_{\beta}f_{\gamma}(q,k_1, k_2),
\end{equation}
where $f_{\nu}(q, k_1, k_2)$ is a function of $q, k_1, k_2$ and
regular at zero momenta.
In the vertex function (\ref{next}),
the effecive vertex $\Gamma_{\mu\alpha\beta\gamma}$ is
contracted with the parity-odd part of the gauge field propagator,
$\Gamma_{\mu}^{(2)}$ of the diagram (a) vanishes.
We therefore find  that the next-to-leading order correction in
$1/N$ to the magnetic moment of anyon is zero in the heavy
anyon limit.
Since there is no correction to the Chern-Simons
term in the $CP^{N-1}$ model \cite{coleman, lee, rajeev},
$g=2$ still holds up to the order of $1/N^2$.
Chou {\it et al.}'s result is therefore verified to this order.

Finally, we consider the two-particle scattering amplitudes
of anyons. The two-particle interaction
due to the gauge particle exchange is shown in figs. 3a, b.
The scattering amplitude for two identical anyons is
\begin{equation}
A_{sym}=A(p_1,p_2;p_1-q,p_2+q)+A(p_1,p_2;p_2+q,p_1-q),
\end{equation}
where
\begin{equation}
A(p_1,p_2;p_1-q,p_2+q)=-i\Gamma_{\mu}(p_1,p_1-q)G_{\mu\nu}(q)
\Gamma_{\nu}(p_2,p_2+q)
\end{equation}
with the vertex function
\begin{equation}
   \Gamma_{\mu}(p_1,p_1-q)=
   (2p_1-q)_{\mu}-i\mu
   {\epsilon_{\mu\nu\lambda}{p_1}_{\nu}q_{\lambda}\over M}.
\end{equation}
For the small momentum transfer, $-q^2\ll M^2$, we get after simple
calculations
\begin{equation}
A_{sym}=4(p_1\cdot p_2)
        {{M\over6\pi}\over
        \left({M\over6\pi}\right)^2+ \left({\theta\over16}\right)^2q^2}
        \left[ 1+2\mu M+O(q^2)\right].
\end{equation}
Therefore we see that for $2\mu M<-1$ or $-{8\over\pi}<\theta<0$
two identical anyons attract each other.

In conclusion, we have calculated the magnetic moment of anyons
in the $CP^{N-1}$ model with a Chern-Simons term in various limits
and found that
the gyromagnetic ratio of anyons is 2 up to the next-to-leading
order in $1/N$ expansion for the heavy anyon limit.
This result is consistent with
the result claimed in \cite{nair}.
We have also shown that for $-{8\over\pi}<\theta<0$
two identical anyons attract each other, which indicates that
anyon and magnetic flux composit is possible.

\vskip .1in
\noindent

{\bf Acknowledgments}
\vskip .1in

This work was supported in part by the Korea Science and Engineering
Foundation through SRC program of SNU-CTP, by NON DIRECTED RESEARCH
FUND, Korea Research Foundation, and also by Basic Science Research
Program, Ministry of Education, 1995 (BSRI-95-2413).

\pagebreak

\section*{Figure Captions}
\typeout{Figure Captions}
\begin{description}
\item[Fig. 1:]
The leading order corrections to the vertex function.
The solid lines denote fermions,  the wavy lines gauge fields, and
The dotted lines the auxiliary fields.

\item[Fig. 2:]The next-to-leading order corrections to the
vertex function.

\item[Fig. 3:] Two-particle interaction due to gauge particle
exchange.

\end{description}
\pagebreak
\begin{figure}
 \epsfysize24cm
 \epsffile[-60 -100 400 800]{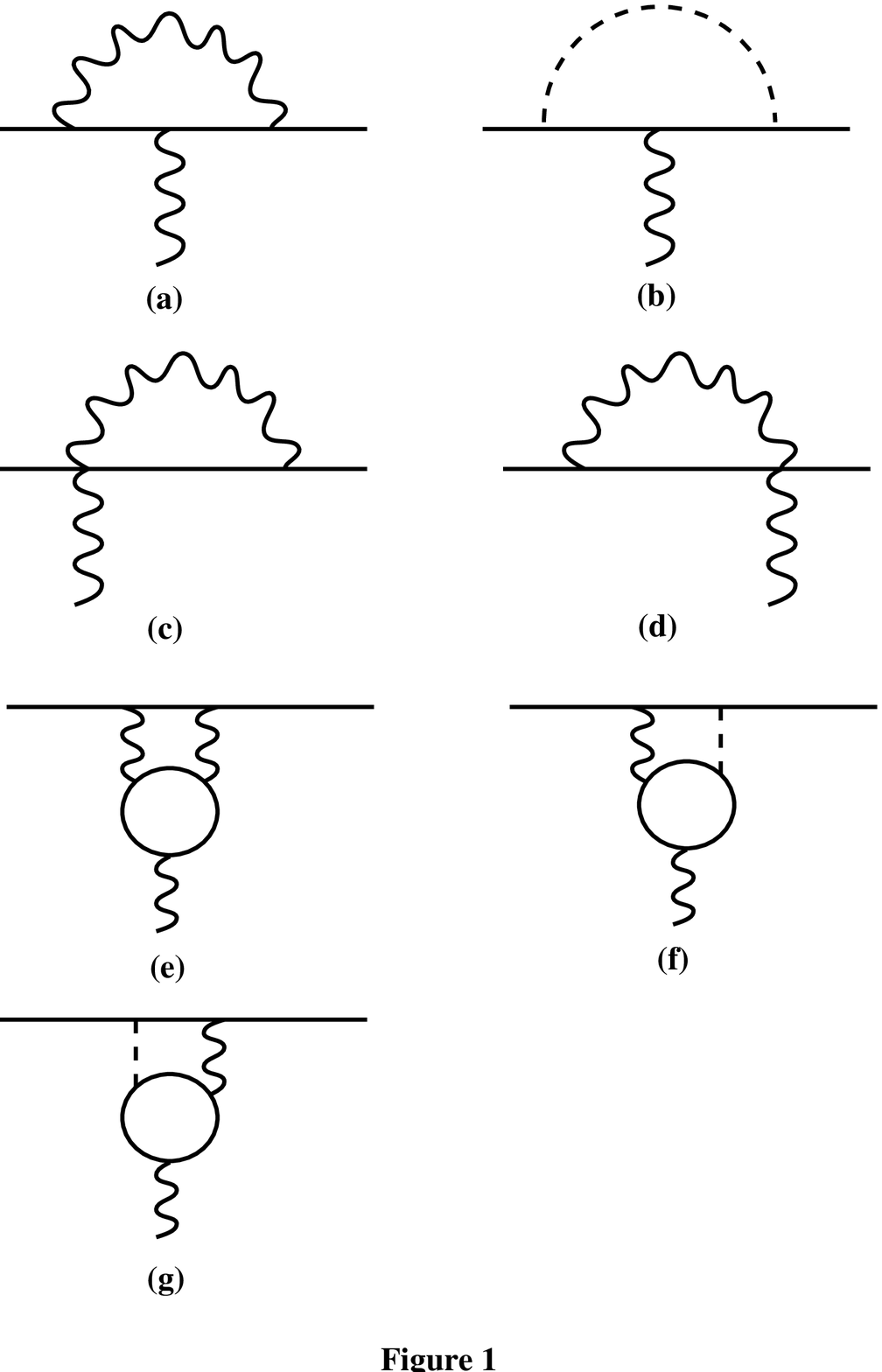}
\end{figure}

\begin{figure}
 \epsfysize24cm
 \epsffile[-60 -100 400 800]{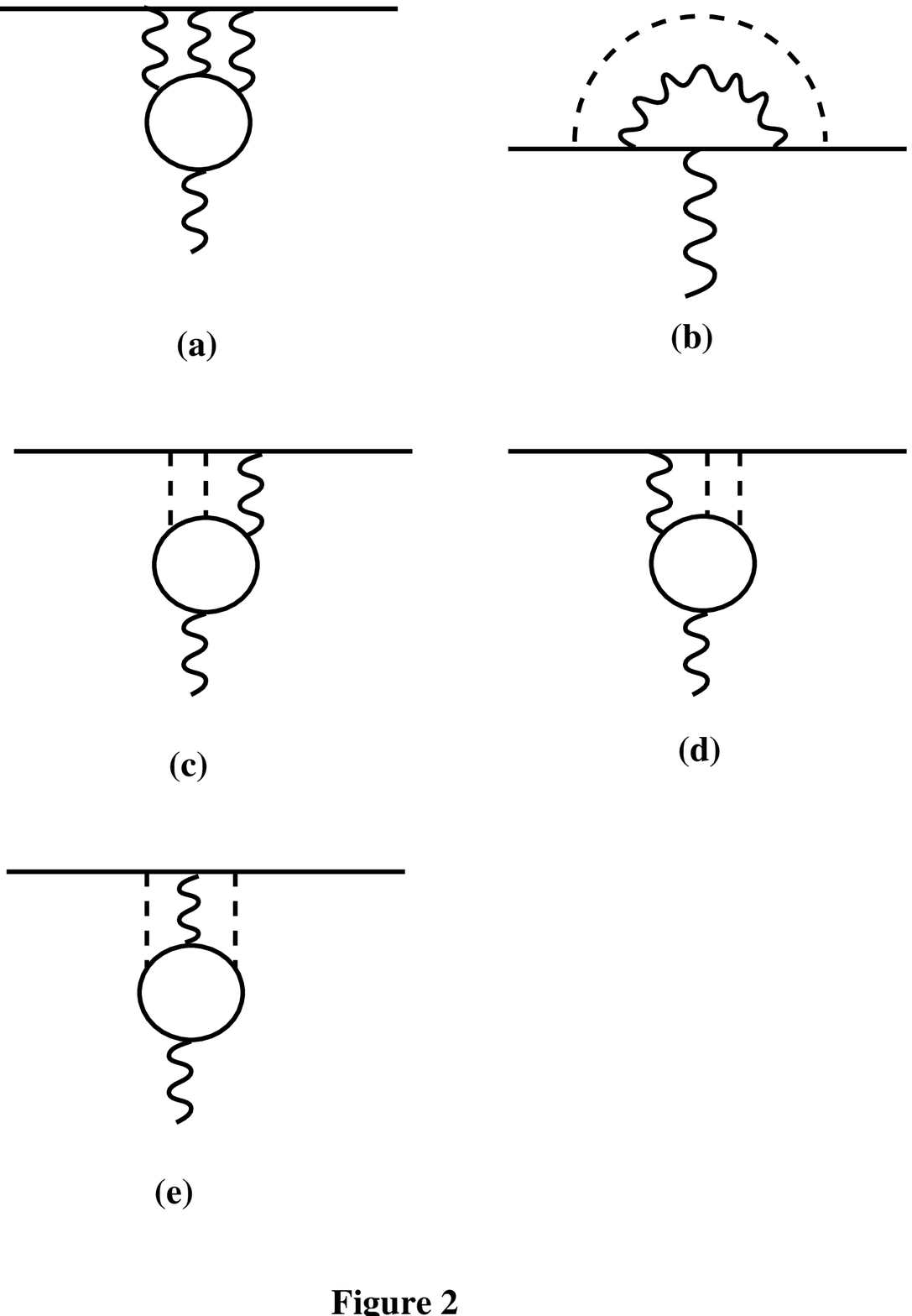}
\end{figure}

\begin{figure}
 \epsfysize24cm
\epsffile[-60 -100 400 800]{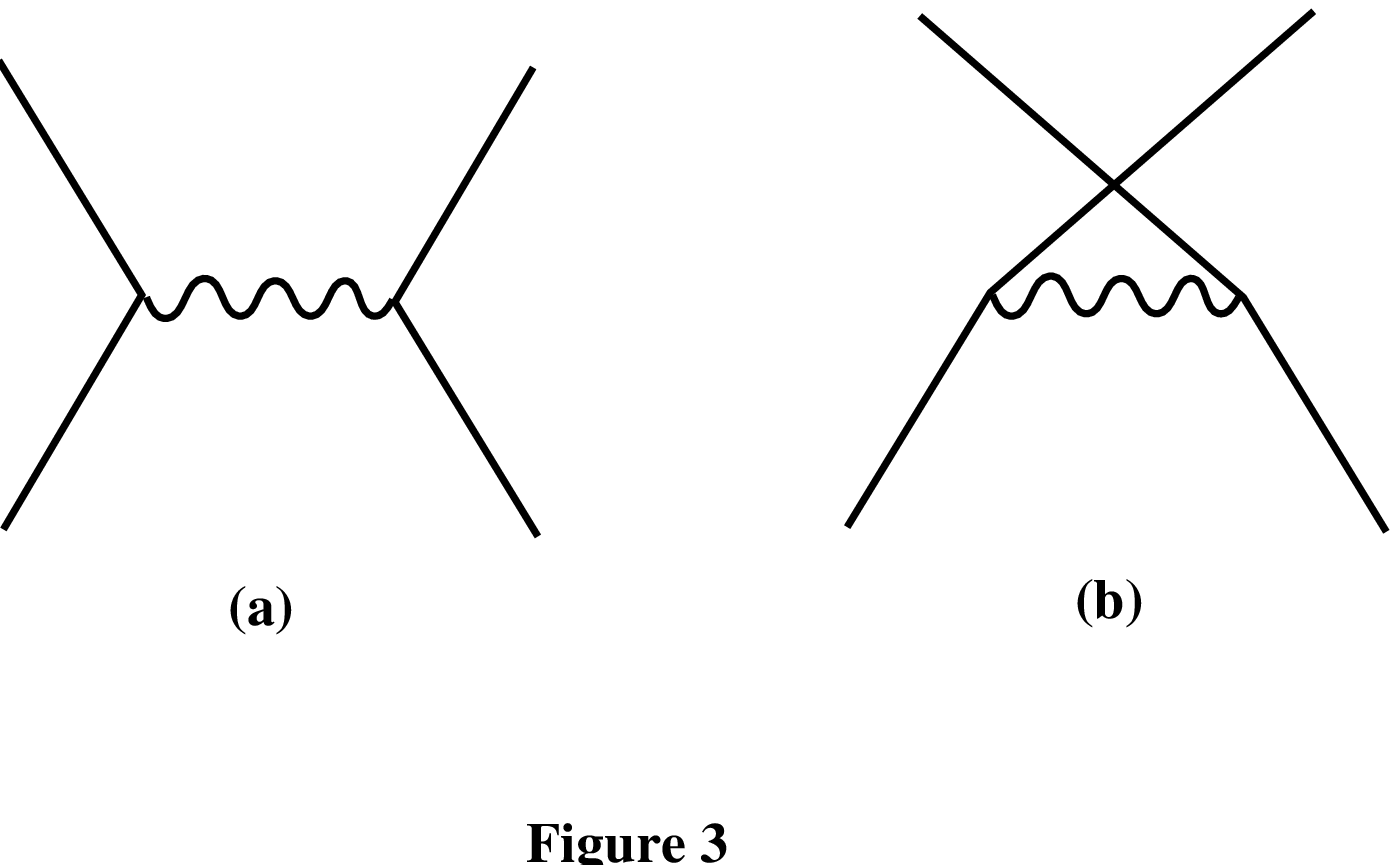}
\end{figure}

\end{document}